\documentclass[twocolumn,english,aps,superscriptaddress, pra,groupedaddress]{revtex4-2}

\usepackage{xcolor}
\usepackage[latin9]{inputenc}
\usepackage{amsmath}
\usepackage{amssymb}
\usepackage{graphicx}
\usepackage{babel}
\usepackage{mathrsfs}
\usepackage{amsfonts}
\usepackage{epstopdf}
\usepackage{multirow}

\begin{document}
\title{Emergent Infrared  Conformal Dynamics in Strongly Interacting Quantum Gases}

\author{Jeff Maki}
\affiliation{Pitaevskii BEC Center, CNR-INO and Dipartimento di Fisica, Universit\`{a} di Trento, I-38123 Trento, Italy}
\author{Fei Zhou}
\affiliation{Department of Physics and Astronomy, University of British Columbia, 6224 Agricultural Road, Vancouver, BC, V6T 1Z1, Canada}

\begin{abstract}
Conformal dynamics can appear in quantum gases when the interactions are fine tuned to be scale symmetric. One well-known example of such a system is a three-dimensional Fermi gas at a Feshbach resonance. In this letter, we illustrate how conformal dynamics can also emerge in the infrared limit in one-dimensional harmonically trapped Fermi gases, even though the system may not have exactly scale symmetric interactions. The conformal dynamics are induced by strong renormalization effects due to the nearby infrared stable scale invariant interaction. When the system approaches the infrared limit, or when the external harmonic trapping frequency $\omega_f \rightarrow 0$, the dynamics are characterized by a unique vanishingly small dissipation rate, $\Gamma \propto \omega_f$, rather than a constant as in generic interacting systems. We also examine the work done in a two-quench protocol, $W$, and the average power $\mathcal{P}$. In one dimension, the average power, $\mathcal{P} \propto \omega_f$, becomes vanishingly small in the infrared limit, a signature of emergent conformal dynamics.

\end{abstract}

\date{\today}
\maketitle

There have been intensive efforts to understand the effects of scale symmetry in strongly interacting Fermi gases \cite{Pitaevskii97, Werner06, Son07, Enss11,  Taylor12, Vogt12, Moroz12, Elliott14, Deng16, Deng18,Maki18, Maki19,Maki20,Maki22, Gritsev10, Bekassy22,Hofmann12}. One remarkable dynamic consequence of scale symmetry is that free space expansion, as well as expansion inside harmonic traps, of strongly interacting Fermi gases can exhibit very unique features which we have denoted as {\em  conformal dynamics}; the unitary evolution can be fully self-similar and entirely equivalent to a simple time-dependent {\em dilation} of the original many-body wavefunction. Microscopically, the hallmark of conformal dynamics is that the $N$-body density matrices are all characterized by functions that are invariant under  conformal transformations \cite{Maki19}.


Normally, the appearance of conformal dynamics requires that the interactions be {\it fully} scale invariant, i.e. the system is residing exactly at a fixed point 
under scale transformation. This typically requires fine tuning the atomic interactions, say, using Feshbach resonances \cite{Chin10}.
One important physical parameter to characterize conformal dynamics and its breaking is the entropy production rate. Conformal symmetry implies that there exists a class of expansion or contraction dynamics with zero entropy production.
This is in stark contrast to the expansion of generically interacting gases which usually results in a finite amount of entropy production and are irreversible \cite{Maki20}.





As far as the entropy production rate is concerned,
any finite deviation from scale symmetric interactions can result in qualitatively different dynamics, depending on whether the interactions are relevant or irrelevant in the long time limit. 
The relevancy of scale breaking interactions on the low-energy physics depends on how these interactions rescale as one approaches long wavelengths or long time scales, i.e. the infrared (IR) limit. 
 If the effects of breaking scale symmetry are IR irrelevant, entropy production can be restricted to short time scales, while the long time dynamics can still be isentropic and governed by {\em an emergent conformal symmetry}. This {\em emergent infrared conformal symmetry} is not associated with the exact symmetry of the microscopic interactions, as in the case of resonance, rather it is an IR symmetry which only appears in the long wavelength or low frequency dynamics. On the other hand, if the scale breaking interactions are IR relevant, one expects an appreciable entropy production rate in the low frequency dynamics. This situation occurs for three-dimensional (3D) spin-$\frac{1}{2}$ Fermi gases near a Feshbach resonance; any minute breaking of scale symmetry in the interactions is amplified in the long-time dynamics due to renormalization effects \cite{Maki18, Maki19}.

The relevancy of scale breaking interactions allows one to address
when conformal dynamics can emerge without fine tuning the interactions. Below we will illustrate such a possibility in one-dimensional (1D) spin-$\frac{1}{2}$ Fermi gases with repulsive interactions near {\em the infrared stable fixed point}. We show that the dynamics can approach scale symmetric ones in the IR limit, even if the microscopic interactions explicitly break scale symmetry, i.e. the dynamics are fully dictated by {\em the infrared stable scale symmetric interactions}. We denote the resulting dynamics as {\em emergent infrared conformal dynamics} ({\it EIRCD}) to distinguish from the exact conformal dynamics ({\it ECD}) that has been discussed previously in the context of $3D$. The main foci of this work are i) the origins of {\em EIRCD}; ii) the similarities and differences between the {\em EIRCD} and {\em ECD}; and iii) the experimental smoking guns of {\em EIRCD}.


{\em i) {\it EIRCD} due to infrared renormalization dynamics:}
In this letter we discuss the possibility of {\em EIRCD} using a model Hamiltonian for a spin-$\frac{1}{2}$ Fermi gas with zero-ranged interactions in $d$ dimensions and in the presence of a time-dependent harmonic trap: $\mathcal{H}_{\omega(t)} = \mathcal{H} + \omega^2(t) C$, where $\omega(t)$ is the time-dependent frequency of the harmonic trap, and $C =\int d^dr \frac{r^2}{2} \psi^{\dagger}({\bf r}) \psi({\bf r})$. We also define $\mathcal{H}$ as the Hamiltonian:
\begin{eqnarray}
\mathcal{H} &= &\int d^d{\bf r} \ \frac{1}{2}\psi^{\dagger}({\bf r}) \left(-\frac{1}{2}\nabla^2\right) \psi({\bf r}) \nonumber \\
&+& \int d^d{\bf r} \ g(\Lambda) \psi^{\dagger}({\bf r})\psi^{\dagger}({\bf r})\psi({\bf r})\psi({\bf r})
\label{eq:Hamiltonian}
\end{eqnarray}
\noindent In Eq.~(\ref{eq:Hamiltonian}), $\psi^\dagger({\bf r})$ is the fermionic creation operator, and  $g(\Lambda)$ is the coupling constant which depends on the ultraviolet (UV) scale of the theory, $\Lambda$. (We have also muted the spin indices and have set $\hbar = m = 1$.)





Since the dynamics are intimately related to the renormalization of the scale symmetry breaking interactions, it is beneficial to briefly recall how the Hamiltonian in Eq.~(\ref{eq:Hamiltonian}) (in units of $\Lambda^2$ {\em so it is dimensionless}) behaves under a scale transformation. We rewrite the interaction as $g(\Lambda)\Lambda^{d-2}=c_d \tilde{g}(\Lambda)$ where $\tilde{g}=\tilde{g}^* +\delta \tilde{g}$ and $c_d$ is a non-zero positive constant that only depends on the dimension $d$. The dimensionless interaction constant $\tilde{g}^*$ represents the scale invariant part of the interactions while $\delta\tilde{g}$ is the part that breaks scale symmetry. 
The value of $\tilde{g}^*$ can be obtained via the standard renormalization group equations (RGEs). Apart from the trivial non-interacting fixed point, $\tilde{g}^* =0$, there is a strongly interacting fixed point $\tilde{g}^*=(2-d)$ when $d=1,3$.
 
We can expand the dimensionless Hamiltonian as $\mathcal{H}=\mathcal{H}^* +\delta \mathcal{H}_I$, where $\mathcal{H}^*$ represents the scale symmetric interacting Hamiltonian while  $\delta \mathcal{H}_I$ 
represents the interactions that break scale symmetry and is proportional to $\delta\tilde{g}(\Lambda)$. Under a scale transformation the residual interaction term $\delta \mathcal{H}_I$ changes according to:
\begin{align}
\Lambda &\rightarrow \frac{\Lambda}{\lambda}, & {\cal \delta H}_I  &\rightarrow \frac{1}{{\lambda}^\eta}  {\cal \delta H}_I 
\label{eq:delta}
\end{align} 
\noindent where $\lambda$ is a scaling factor set to be larger than unity for our discussions, i.e. $\lambda >1$, as this generates a flow towards the IR limit. The scaling dimension of ${\cal \delta H}_I$, $\eta=\eta(d, \tilde{g}^*)$, is universal only depending on the spatial dimension $d$ and $\tilde{g}^*$. It can again be directly obtained by solving the RGEs around $\tilde{g}^*$.

In 3D or $d=3$, there is a scale invariant attractive interaction, $\tilde{g}^*= - 1$, which has been identified as a Feshbach resonance with an infinite scattering length $a_{3D} =+\infty$ \cite{Sachdev}. The scaling dimension for $\delta \mathcal{H}_I$ is $\eta= -1 <0 $. The scale invariant fixed point Hamiltonian ${\cal H}^*$ is IR unstable under a scale transformation; any small deviations of $\delta \mathcal{H}_I$ from the scale invariant fixed point will be amplified during the course of rescaling towards the IR limit \cite{Maki19}. 
In 1D, the scale invariant interactions are repulsive, $\tilde{g}^*=+1$, and represent infinite strength contact interactions, i.e. the 1D scattering length $a_{1D}=0$. The associated scaling dimension is $\eta=+1 >0$,
and the effect of ${\cal \delta H}_I$ diminishes as the running scale $\Lambda$ is lowered (or  equivalently the scaling factor $\lambda$ increases); $\mathcal{H}^*$ is IR stable. 

In the quantum dynamics of interest, the particle density $n(t)$ and the characteristic chemical potentials $\mu(t)  \sim n^{2/d}(t)$ depend on time $t$ and can dramatically differ from their initial values at $t=0$. We quantify the effects of ${\cal \delta H}_I$ on the dynamics by evaluating it at the energy scale $\mu(t)$, or equivalently at a momentum scale set by the instantaneous Fermi momentum: $k_F(t) \propto n^{1/d}(t)$. It is then natural to define a time dependent running scale at which we evaluate $\mathcal{\delta H}_I$ that is proportional to the ratio of the density at time $t$ to the initial density: $\Lambda(t) \approx \left(n(t)/n(0)\right)^{1/d} \Lambda(0)$. In this way we can define a dynamic rescaling factor $\lambda(t) \sim \left(n(t)/n(0)\right)^{1/d}$. 




From this point of view, expansion or contraction dynamics can be studied by examining the renormalization flow of the symmetry breaking action $\delta \mathcal{H}_I$ \cite{RG_note}.
In order to observe substantial renormalization effects during the course of the dynamics, it is thus imperative to consider the far-away-from-equilibrium expansion (contraction) dynamics, 
so that the rescaling factor $\lambda(t) $ can be much larger (smaller) than unity. This is contrast to the perturbative linear response regime where $\lambda(t)$ always remains close to one.

One can then directly study {\em ECD} and the possibility of {\em EIRCD} by considering a thermally equilibrated quantum gas that is released from a very tight isotropic $d$-dimensional harmonic trap of frequency $\omega_i$ into a much softer harmonic trap with frequency $\omega_f$. Provided $\omega_f \ll \omega_i$, one can then generate an appreciable dynamic renormalization group flow towards the IR limit. 

When the interactions are exactly scale symmetric, there are {\em ECD}. One can show that the scale parameter for this quench protocol is given by \cite{Werner06,Moroz12, Maki20}:
\begin{align}
\lambda^2(t) &= \cos^2(\omega_f t) + \frac{1}{\tilde{\omega}_f^2} \sin^2(\omega_f t), &\tilde{\omega}_f&=\frac{\omega_f}{\omega_i}.
\label{eq:lambda_t}
\end{align} 
\noindent  Eq.~(\ref{eq:lambda_t}) is a periodic function with period $T_f = \pi/\omega_f$. It takes on values: $\lambda(t) \in [1,\lambda_{max}]$, with a maximum value: $\lambda_{max}=1/\tilde{\omega}_f$ and a minimum value: $\lambda_{min} = 1$. 

To further illustrate this limit, we emphasize an important signature of {\em ECD}, the dynamics of the Wigner function and the momentum distribution. The Wigner function is related to the one-body density matrix $\rho_1({\bf r}, {\bf r}'; t)$ via:

\begin{equation}
\rho({\bf R},{\bf k};t) =\int d^d{\bf r} e^{-i {\bf k}\cdot{\bf r}} \rho_1\left(\frac{{\bf R}}{2} +{\bf r}, \frac{{\bf R}}{2}-{\bf r};t\right)
\end{equation}  where ${\bf R}=({\bf r}+{\bf r}')/2$ is the center of mass coordinate and  ${\bf k}$ is the relative momentum. 
The conformal symmetry heavily constrains the dynamics of the one-body density matrix to the following form \cite{Maki19}:

\begin{align}
\rho_1( {\bf r},{\bf r}', t) &= \frac{1}{\lambda^{d}(t)} \exp\left[i \frac{\dot{\lambda}(t)}{\lambda(t)} \frac{r^2-r'^2}{2}\right]  \rho_1\left(\frac{{\bf r}}{\lambda(t)},\frac{ {\bf r}'}{\lambda(t)} , 0\right)\nonumber \\
\label{app:dm_final}
\end{align}

\noindent where $\lambda(t)$ is defined in Eq.~(\ref{eq:lambda_t}). Eq.~(\ref{app:dm_final}) indicates that the Wigner function has the property:

\begin{equation}
 \rho ({\bf R}, {\bf k};t) = \rho\left(\frac{{\bf R}}{\lambda(t)}, {\bf k} \lambda(t) - \frac{\dot \lambda(t)}{\lambda(t)} {\bf R}; 0\right)
\label{eq:Wigner}
\end{equation}

\begin{figure}
\includegraphics[scale=0.55]{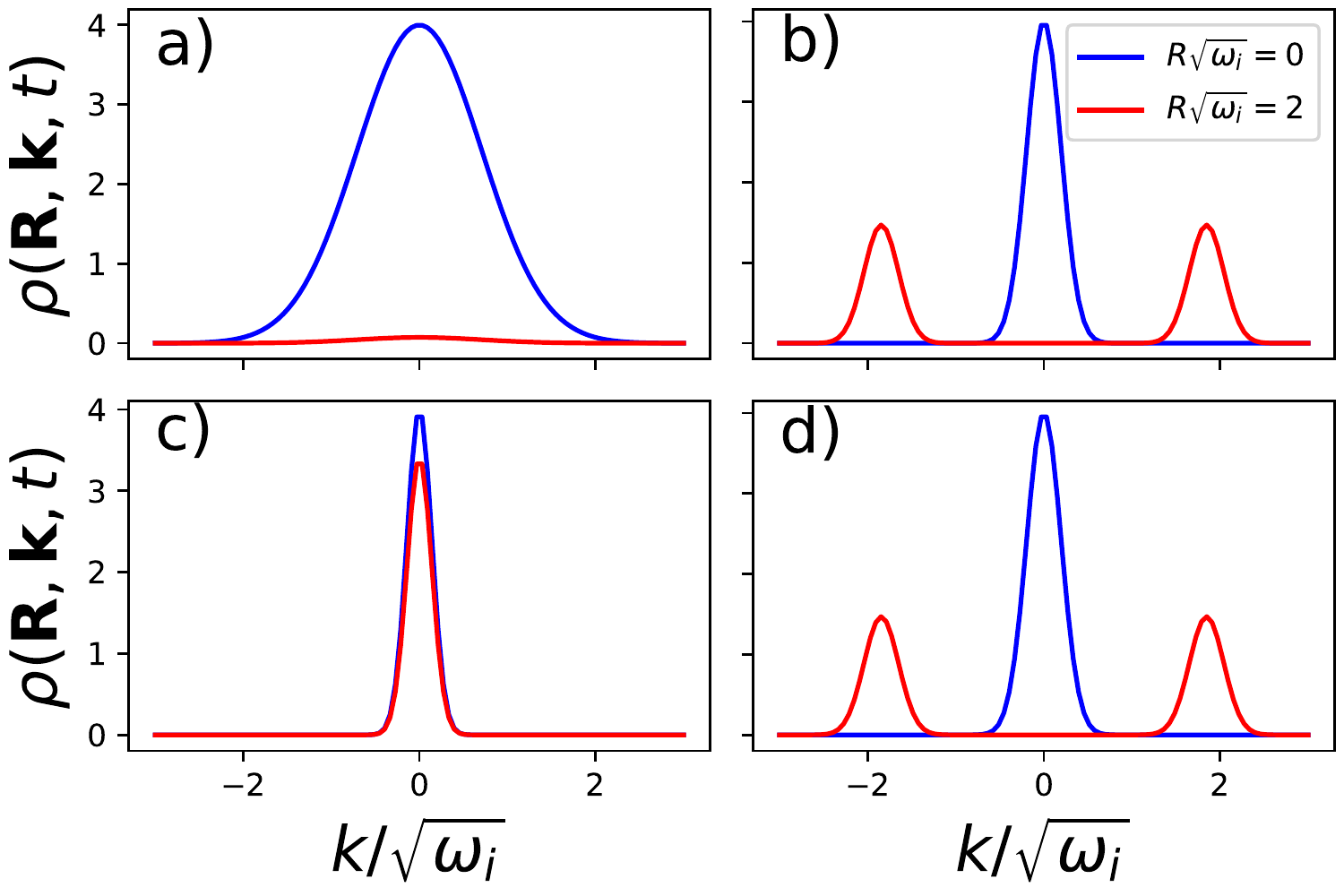}
\caption{Schematic for the Wigner distribution, Eq.~(\ref{eq:Wigner}), as a function of momentum $k$ for two center-of-mass coordinates, $R$ (blue: $R=0$, and red: $R = 2/\sqrt{\omega_i}$). The plots $a)-d)$ correspond to times $ t/T_f = 0, 1/4, 1/2, 3/4$, respectively. When $R=0$ the dynamics are a simple rescaling, while for finite $R$ the dynamics also  involves a time-dependent shift of the center of the Wigner distribution.} 
\label{fig:wigner}
\end{figure}

The dynamics of Eq.~(\ref{eq:Wigner}) are schematically shown in Fig.~(\ref{fig:wigner}) for fixed ${\bf R}$. The ${\bf k}$ dependence of the Wigner function dilates with time according to $\lambda(t)$, while the center of the function shifts by an amount ${\bf R}\dot{\lambda}(t)/\lambda(t)$. Such dynamics are again completely periodic with a period $T_f$. A natural corollary of Eq.~(\ref{eq:Wigner}) is that the momentum distribution:
$n(k,t) = \int d^d{\bf R} \rho({\bf R}, {\bf k},t)$ is completely oscillatory in {\em ECD} with a period $T_f$ .  

These oscillations were originally posited in Ref. \cite{Minguzzi05} and seen experimentally \cite{Wilson20} for 1D Bose gases in the Tonks-Girardeau (TG) limit. Here we state that this phenomenon is more generally associated with the {\em ECD} of scale invariant quantum gases expanding in harmonic traps, and thus can be observed in higher-dimensions, for example the 3D spin-$\frac{1}{2}$ Fermi gases at a Feshbach resonance. 



We can examine the effects of $\delta \mathcal{H}_I$ on the conformal dynamics by treating $\delta{\mathcal{H}}_I$ perturbatively.
At leading order, the scale breaking perturbation effectively becomes time dependent during the course of the dynamics. We can capture the time-dependence by rescaling $\mathcal{\delta H}_I(t)$ by $\lambda(t)$ defined in Eq.~(\ref{eq:lambda_t}). Thus the time-dependence of the scale breaking perturbation sensitively depends on the scaling dimension $\eta$ introduced in Eq.~(\ref{eq:delta}).
In 3D,  $\eta=-1$ and $\delta \mathcal{H}_I(t)$ is appreciable over the entire period and becomes more relevant when $\lambda(t)$ approaches $\lambda_{max}$, i.e. when $t \simeq T_f/2$. 
In 1D, $\eta=+1$ and $\mathcal{\delta H}_I(t)$ is only appreciable for small time windows $\delta t \sim 1/\omega_i$ near $t=0$ or $t=T_f$  when the scaling factor $\lambda (t) \sim 1$ is at a minimum and the gas is most dense; $\mathcal{\delta H}_I$ is strongly suppressed around $t=T_f/2$ when the gas is most dilute, $\lambda(t) \approx \lambda_{max}$.
Accordingly, there are {\em EIRCD} in 1D, provided $\tilde{\omega}_f$ is sufficiently small, while in 3D {\em EIRCD} are absent.

{\em ii) EIRCD vs ECD:} 
When the interactions break scale symmetry, the oscillatory conformal dynamics in Eq.~(\ref{eq:lambda_t}) will become damped, a signal of entropy production. 
We quantify the extent of the  breaking of scale symmetry and the onset of the {\it EIRCD} using $\Gamma (\tilde{\omega}_f)$, the decay rate of the oscillatory dynamics averaged over the period.
When $\tilde{\omega}_f \rightarrow 0$,
we can expand the decay rate in terms of $\tilde{\omega}_f$,
\begin{eqnarray}
{\Gamma} &=& \Gamma_0 +\Gamma_1 \tilde{\omega}_f+\Gamma_2 \tilde{\omega}_f^2+...; \nonumber \\
\Gamma_{0,1,2} &=& \omega^2_i \tau_F  \gamma_{0,1,2} \left[\left(\frac{a_{sc}^2}{\tau_F}\right)^{2-d}, T\tau_F, \omega_i \tau_F \right]
\label{eq:rate}
\end{eqnarray}
where $\tau_F \propto n(0)^{-2/d}$ is a Fermi time defined in terms of the initial density at the center of the trap, and $\gamma_{0,1,2}$ are dimensionless functions. 
The general structure of the coefficients in Eq.~(\ref{eq:rate}) is highly complex and we will focus on the limit $(a_{sc}^2/\tau_F)^{2-d} \ll 1$, with $a_{sc}$ being the $d$-dimensional scattering length: $a_{sc}=a_{3D}$ for $d=3$ and $a_{sc} =a_{1D}$ for $d=1$. 

When ${\mathcal{H}}$ is fine tuned to have scale invariant interactions, i.e. $\mathcal{H} = {\mathcal{H}}^*$, $\Gamma = 0$ for all values of $\tilde{\omega}_f$ and arbitrary temperatures $T$ \cite{Maki19, Maki20}. This is a signature of {\it ECD} and is also fully consistent with the absence of bulk viscosity in the high temperature thermal phase \cite{Son07}.
For generically interacting systems ${\Gamma}$ will be finite, and will be proportional to $(a_{sc}^2/\tau_F)^{2-d}$ near the strongly interacting fixed point. However, the dissipation in the IR limit  sensitively depends on whether $\gamma_{0}$ or $\gamma_{1,2,...}$ is the leading contribution to the decay rate.

The leading behaviour of ${\Gamma}$ for finite $\mathcal{\delta H}_I(t)$ in the IR limit is also directly related to the value and sign of the scaling dimension $\eta$ defined in Eq.~(\ref{eq:delta}). 
When $\eta=-1$, as in 3D Fermi gases near resonance, the symmetry breaking action of $\delta \mathcal{H}_I(t)$ is relevant as the gas expands.
The dissipation occurs approximately uniformly over the whole period $T_f$ leading to an $\omega_f$-independent time-averaged dissipation rate, or a finite $\Gamma_0$(see detailed discussions below). Thus ${\Gamma} \rightarrow \omega^2_0 \tau_F \gamma_0$ is approximately a constant as $\tilde{\omega}_f \rightarrow 0$, as in generic strongly interacting systems. This signifies finite dissipation in the IR limit and the absence of {\em EIRCD}.

\begin{figure}
\includegraphics[scale=0.50]{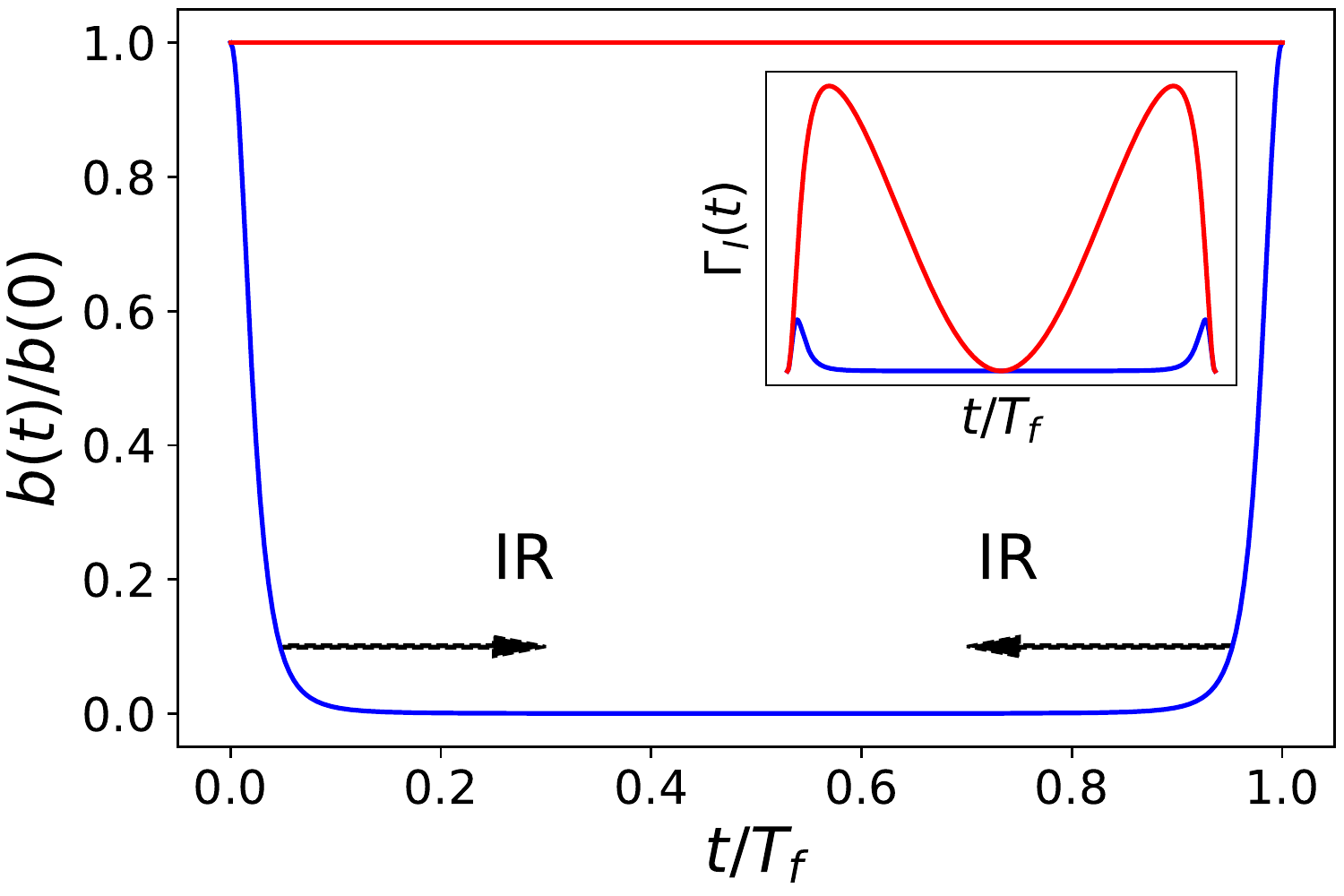}
\caption{The time-dependent damping coefficient $b(t)$, Eq.~(\ref{eq:diss_hydro}), as a function of time for one period of the conformal dynamics in both 1D (blue) and 3D (red). In 1D the dissipation is only appreciable near $t=0, T_f$ when the gas $\lambda(t)\sim 1$, while in 3D there is appreciable dissipation over the entire period. The arrows points towards the IR limit in the dynamics where the scaling factor $\lambda(t=T_f/2) \rightarrow \lambda_{max}$. The inset shows the instantaneous damping rate, $\Gamma_I(t)$, over the same time range.}
\label{fig:gamma}
\end{figure}
By contrast,
when $\eta=1$, as in 1D, $\delta \mathcal{H}_I(t)$ becomes strongly suppressed in the bulk of the period, when $\lambda(t)  \gg 1$. The dissipation due to $\delta \mathcal{H}_I(t)$ is then confined to a small time window of $\delta t \sim \pi/\omega_i \ll T_f $ when $\lambda(t)\approx1$, i.e. $t=0, T_f, 2 T_f$ etc. The averaged dissipation rate therefore is inversely proportional to $T_f$ or proportional to $\omega_f$. So we expect that $\gamma_{0}=0$ while $\gamma_1$ remains finite. This scaling implies that if one approaches the IR limit, $\tilde{\omega}_f \rightarrow 0$, ${\Gamma}$ becomes vanishingly small indicating an unexpected dynamical phase with {\it EIRCD}. 
Such nearly dissipationless dynamics can be directly studied in experiments.

To provide more evidence for this phenomenology, we now restrict ourselves to the hydrodynamic limit and evaluate the instantaneous dissipation rate, $\Gamma_I(t)$, to leading order in the scale breaking interactions. 
We solve the Navier-Stokes equation \cite{Landau_Fluid} for a $d$-dimensional gas in a harmonic trapping potential:

\begin{equation}
n\left(\partial_t +  {\bf v} \cdot \nabla_{\bf r} \right) {\bf v} = -\nabla_{\bf r} P - n \omega_f^2 {\bf r}_i + \nabla_{\bf r} \left(\zeta \nabla_{\bf r} \cdot {\bf v}\right)
\label{eq:NS}
\end{equation}

\noindent 
$n$, ${\bf v}$ and $P$ are the local density, velocity and pressure, respectively. The last term in Eq.~(\ref{eq:NS}) is the dissipative term which defines the bulk viscosity, $\zeta$ (we have muted the shear viscosity term as it is not needed for the breathing dynamics).
At the scale invariant fixed point,
Eq.~\eqref{eq:NS} can be solved using
a scaling ansatz for the density $n({\bf r},t) = \lambda^{-d}(t) n\left({\bf r}/\lambda(t),0\right)$ and for the velocity field: ${\bf v}({\bf r},t) = {\bf r} \dot{\lambda}(t)/\lambda(t)$, where $\lambda(t)$ is again given by Eq.~\eqref{eq:lambda_t}. 
This ansatz is consistent with the Wigner distribution function, Eq.~(\ref{eq:Wigner}), and can also be obtained microscopically by
applying an $SO(2,1)$ conformal-field-theory to the dynamics at the scale invariant fixed point \cite{Maki20, Maki22}. 
The dynamics when $\delta \mathcal{H}_I=0$ are again non-dissipative, while for finite $\delta \mathcal{H}_I$, the main effects of the broken scale symmetry are captured by the finite bulk viscosity.

In this hydrodynamic framework the instantaneous dissipation rate is defined as $\Gamma_I(t) = \left|\frac{1}{E} \frac{d\langle E_k\rangle(t)}{dt}\right|$ where $E$ is the total energy per particle
and $\langle E_k \rangle(t)$
is the motional energy associated with the hydrodynamic flow per particle. From Eq.~(\ref{eq:NS}) one can then show that the instantaneous dissipation rate is itself proportional to the bulk viscosity: $d\langle E_{k}\rangle/dt = -d \int \frac{d^d {\bf r}}{N} \zeta({\bf r},t)  \left(\frac{\dot{\lambda}(t)}{\lambda(t)}\right)^2$. Near the scale invariant fixed-point the bulk viscosity possesses a simple scaling form \cite{Enss19, Nishida19, Hofmann20} which allows one to write:
\begin{align}
\Gamma_I(t) &= b(t) \dot{\lambda}^2(t),  & b(t)&= \frac{e_d}{\omega_i \lambda^2(t)} \left(\frac{k_F a_{sc}}{\lambda(t)}\right)^{2\eta}.
\label{eq:diss_hydro}
\end{align}

\noindent In Eq.~(\ref{eq:diss_hydro}), $\eta=2-d$ for $d=1,3$ and $e_d$ is a time-independent constant that depends on the equation of state of the initial gas. $k_F$ is the Fermi wavenumber defined at the center of the trap ($k_F = \sqrt{2/\tau_F} \propto n(0)^{1/d}$), $a_{sc}$ is again the $d$-dimensional scattering length, and $\lambda(t)$ is defined in Eq.~(\ref{eq:lambda_t}).

The time dependence of the damping coefficient $b(t)$ over one period of the dynamics is shown in Fig.~(\ref{fig:gamma}) near the strongly interacting scale invariant point in a) 1D and b) in 3D. In 1D the damping coefficient is strongly suppressed save for two tiny time windows near $t=0$ or $t=T_f$, while in 3D it remains close to unity over the whole period. Taking the average of $\Gamma_I(t)$ over the period $T_f$, we
verify the prediction that the leading term in Eq.~(\ref{eq:rate}) is $\gamma_0$ in 3D, while it is  $\gamma_1$ in 1D.





\begin{figure}
\includegraphics[scale=0.5]{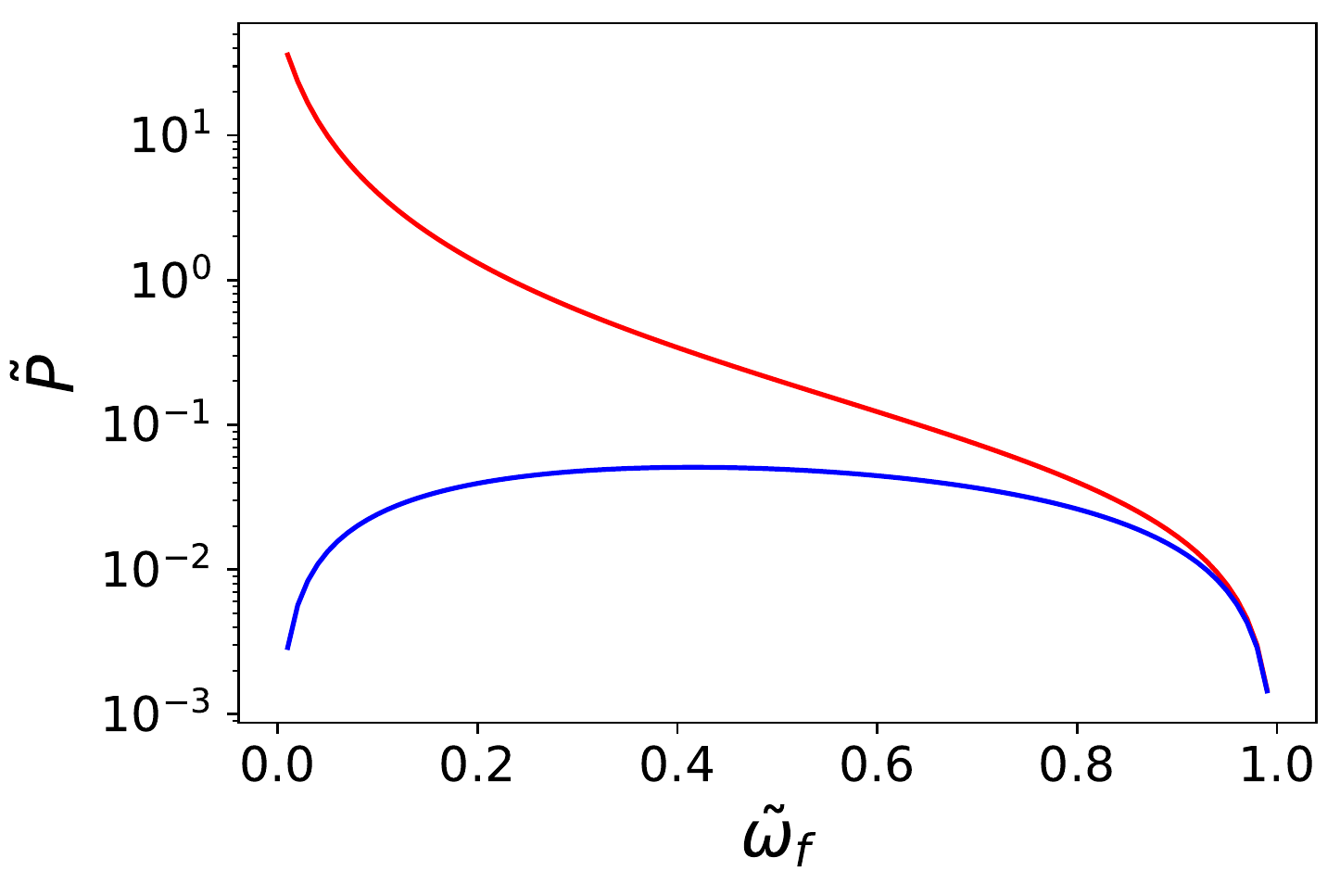}
\caption{Dynamic simulation of the average power of the work done (see the main text): $\tilde{P}$ as a function of $\tilde{\omega}_f$ in 3D (red) and 1D (blue) for dimensionless bulk viscosity $\tilde{\zeta} = d^2 \int d^d{\bf r} \zeta({\bf r},0) /(2N \omega_i \langle C \rangle(0)) = 0.1$.  In the IR limit, $\tilde{\omega_f} \to 0$, the average power, proportional to $\omega_f$, is vanishingly small in 1D, while it diverges as $\omega^{-1}_f$ in 3D. 
}
\label{fig:work}
\end{figure}

{\em iii) A measurement scheme:} 
When the system has {\em EIRCD}, the dynamics are nearly reversible as the entropy production is localized to short time windows near $\lambda(t) = 1$.
A convenient way to experimentally study the reversibility of the dynamics is to examine the work done following a second quench which returns the system to a harmonic trap with frequency $\omega_i$ after hold time $t_h$.

The total work done in this two quench protocol, one quench at $t=0$ and one at $t=t_h$, sensitively depends on the value of $t_h$ and whether there are {\em EIRCD}. Generally, the work done is given by: 
\begin{align}
W&=\frac{(\omega_i^2-\omega_f^2)}{2} \left( \lambda^2_C (t_h) -1 \right) \langle C \rangle (0), & \lambda^2_C(t_h) &=\frac{ \langle C\rangle(t_h)}{\langle C \rangle(0)}.\nonumber\\
\label{eq:work}
\end{align}
\noindent Here we  have introduced  $\lambda_C(t_h)$ as a measure of the actual size of the quantum gas at $t=t_h$, which in general differs from $\lambda(t_h)$ defined in Eq.~(\ref{eq:lambda_t}), but coincides with $\lambda(t_h)$ for {\em ECD}, i.e when $\delta \mathcal {H}_I=0$.  $W$ in this two-quench protocol is positive semi-definite, i.e. $W \geq 0$ for all values of $\omega_i$ and $\omega_f$. 

When $\omega_f \ll \omega_i$, the work done reaches a minimum, $W=0$, if $\lambda_C^2(t_h) =1$. 
This condition can only be satisfied in {\em ECD} when $t_h=nT_f$ $n=1,2,3...$, as $\lambda^2_C(t_h) = \lambda^2(t_h)=1$ \cite{work}.
If there is dissipation, the amount of work done at any hold time $t_h>0$ {\em shall always be larger} than zero, $W(t_h>0)>0$, as $\lambda_C(t_h)$ now has to be larger than its initial value of unity due to entropy production.
There will naturally be residual dynamics for $t > t_h$, even when the holding time is exactly $t_h=n T_f$, unlike in {\em ECD} \cite{Tf}. Thus $W(t_h=T_f)$ being zero or not can be a powerful criterion for detecting {\em ECD} and its breaking. 

The amount of work done after one cycle in this two-quench scheme can also be used as an effective probe of {\it EIRCD}. 
Near a 3D Feshbach resonance, the action of $\delta \mathcal{H}_I(t)$ is enhanced in the IR limit, leading to dissipation or conventional thermalization as $\omega_f$ approaches zero.
So in the IR limit when $\omega_f \rightarrow 0$, the scaling parameter approaches an  equilibrium value:  $\lambda^2_C(t_h=T_f) \rightarrow \lambda_{eq}^2= 1/(2 \tilde{\omega}_f^2)$ which is much bigger than $\lambda^2(t=T_f)=1$. 
This results in $W \propto \tilde{\omega}_f^{-2}$, which diverges as $\tilde{\omega}_f$ approaches zero.

On the other hand in 1D near the strong coupling fixed point, $\delta \mathcal{H}_I(t)$ becomes irrelevant in the IR limit, strongly suppressing entropy production and thermalization. 
Because of the EIRCD, the small amount of dissipation occurs during a short-time window, $\delta t  \sim 1/\omega_i$, that is parametrically small compared to $T_f$.
This indicates $\lambda^2_C(t_h=T_f)=1+ O( \Gamma_1/ \omega_i)$ which is very close to unity as $\Gamma_1/\omega_i \sim \omega_i \tau_F \ll 1$ in the many-body limit. The work done will be small and independent of $\tilde{\omega}_f$ in the IR limit \cite{work2}.



A more convenient way to visualize this physics is to consider the average power: $\tilde{P} = W \omega_f$. Following our previous discussions on the work, $\tilde{P}$ is proportional to $\tilde{\omega}_f$  in 1D and becomes vanishingly small in the IR limit, due to the irrelevancy of $\delta \mathcal{H}_I(t)$. While in 3D the average power diverges as $\tilde{P} \propto 1/\tilde{\omega}_f$. 
We have numerically confirmed this behaviour for the average power and for the work,  by solving the Navier-Stokes equation for trapped gases, Eq.~(\ref{eq:NS}),  using the fore mentioned scaling solution  \cite{Maki20,Maki22}.  
The results of the simulation are presented in Fig.~(\ref{fig:work}), which shows that the average power $\tilde{P}$ as a function of $\tilde{\omega}_f$ is consistent with our scaling analysis.



Before concluding, we want to make two remarks. First, we have parametrized this {\em EIRCD} using the dissipation rate, Eq.~(\ref{eq:rate}). 
We illustrate that in 1D the leading contribution to the decay rate is $\Gamma \propto \tilde{\omega}_f$ as $\tilde{\omega}_f \to 0$ while in 3D $\Gamma$ is approximately constant. This difference controls whether there are {\em EIRCD} or not.
Eq.~(\ref{eq:rate}) ought to be contrasted to {\em linearized hydrodynamics} when  $\lambda(t) =1+\delta \lambda(t)$ with $|\delta \lambda(t)| \ll 1$. This limit is opposite to the deep scaling regime we have focused on in this letter where $\lambda_{max} \gg 1$.
In the IR scaling regime ($\tilde{\omega}_f \rightarrow 0$), the dynamics are highly non-linear but are self-similar and conformal. 
The linearized hydrodynamic limit can be realized when $\tilde{\omega}_f \approx 1$ or $|{\omega}_f -\omega_i|  \ll \omega_i$;
and the damping is given by $\Gamma_{c.m} \approx \omega^2_i \tau_R$, for a relaxation time $\tau_R$. 
Extending our results to the linear response limit naturally reproduces the previous literature on collective modes \cite{BEC_REV,BEC_BOOK}.

Second, our analysis so far has been focused on the spin-$\frac{1}{2}$ Fermi gas. When applied to 1D interacting Bose gases, our theory indicates that {\em EIRCD} shall occur near the strong coupling fixed point, $g^*=+1$ with $\eta=1$, which now can be identified as  the hardcore limit of the Lieb-Liniger model. However, {\em EIRCD} does not occur in weakly interacting Bose  gases near the trivial non-interacting fixed point ($g^*=0$ and $\eta=-1$).


The application of {\em EIRCD} to 1D Bose gases offers a complementary view to the recent numerical studies using generalized hydrodynamics (GHD) \cite{Castro16, Bertini16,Bulchandani17,Caux19, Bouchoule22, Essler22, DeNardis18, Bastianello22}. GHD  takes into account the integrability of the Lieb-Liniger model in free space, and has been successful in describing the motion beyond conventional hydrodynamics, as observed in experiments \cite{Kinoshita04, Moller21, Cataldini22, Schemmer19, Wilson20}. In the presence of an external harmonic trapping potential, integrability is broken leading to thermalization towards the standard Gibbs ensemble \cite{Bastianello20, Thomas21} when the interaction strength is finite. Numerical studies have already suggested that the thermalization rate is an order of magnitude larger in the weakly interacting gases, in comparison to the strongly interacting hardcore limit \cite{Thomas21}. Such a conclusion emerges naturally in our renormalization group analysis, as 1D weakly-interacting gases do not exhibit {\em EIRCD}, as they are represented by an {\it infrared unstable scale invariant fixed point} \cite{Maki18,Maki19}.
An analysis similar to the one above indicates that the dynamics near the 1D non-interacting fixed point are similar to strongly damped dynamics near 3D Feshbach resonances \cite{GHD}.

In summary we have investigated the dynamics of strongly interacting quantum gases released from tight harmonic traps in $d=1,3$-dimensions into much shallower traps. We have found that near the strongly interacting scale invariant point in 1D, the expansion dynamics generates a renormalization group flow towards the IR limit which renders the breaking of scale invariance irrelevant. Since the strongly interacting fixed point in 1D is IR stable and robust, {\em EIRCD} can be viewed as a dynamical phase that remains to be further studied in experiments. It also remains to further explore in a more quantitative manner the relation between the point of view of {\it EIRCD} in the proximity of {\it infrared stable fixed points}, which generally can occur in both 1D and higher spatial dimensions, and fascinating microscopic dynamics of one-dimensional integrable systems such as the Lieb-Liniger model but in a harmonic trap.

{\it Acknowledgements}: We want to thank Randy Hulet, Kirk Madison and Riley Stewart for discussions on possible experimental detections, and Alvise Bastianello for discussions on generalized hydrodynamics.




\begin{thebibliography}{References}
\bibitem{RG_note} In general expansion dynamics contain two parts. The first can be described by the RG flow. The RG flow defines the instantaneous Hamiltonian at time, $t$. There are also non-trivial dynamics within the instantaneous Hilbert space which is beyond the RG flow.
\bibitem{Tf}The period of motion can be slightly shifted away from $T_f=\pi/\omega_f$ in {\em EIRCD} when the interactions are not exactly fine tuned. In experimental studies of $W$, $t_h$ can be adjusted so that $W(t_h)$ reaches a minimum value. In this way one can match $t_h$ to the renormalized value of the period $T^{R}_f$, which will be in the vicinity of $\pi/\omega_f$.
\bibitem{work2}A more experimentally convenient way to extract $W$ is perhaps to examine the size of the gas at $t$ much later than $t_h$, when the isolated quantum gas is fully thermalized in the original trapping potential with frequency $\omega_i$. Taking into account the work-energy relation, for the measured final  and initial sizes  $\langle C \rangle (t=\infty, 0)$, one can infer:
${W}= \omega^2_i \langle  C \rangle(0) \left( \frac{\langle C \rangle (\infty)}{\langle C \rangle (0)}-1 \right)$.
It provides an explicit and alternative relation between $W$ defined in Eq.~(\ref{eq:work}), the work done in the two-quench scheme, and physical observables, $\langle C \rangle (t=\infty, 0)$.
\end{thebibliography}
\end{document}